\documentclass[journal,a4paper]{IEEEtran}

\usepackage[utf8]{inputenc}
\usepackage[cmex10]{amsmath}
\usepackage{graphicx}
\usepackage{amssymb}
\usepackage{amsthm}
\usepackage{subfigure}
\usepackage{cite}
\usepackage{color}
\usepackage{bm}
\usepackage{dblfloatfix}

\usepackage[hidelinks]{hyperref}
\usepackage[author={M. Kokshoorn}]{pdfcomment}
\usepackage{xcolor}

\usepackage{breakurl} 

\theoremstyle{remark}

\usepackage[ruled,norelsize]{algorithm2e}
\makeatletter
\newcommand{\removelatexerror}{\let\@latex@error\@gobble}
\makeatother
\usepackage{algorithmicx}

\PassOptionsToPackage{hyphens}{url}\usepackage{hyperref}

\begin{document}

\title{Fountain Code-Inspired Channel Estimation for Multi-user Millimeter Wave MIMO Systems}
\IEEEoverridecommandlockouts 
\author{
    \IEEEauthorblockN{Matthew Kokshoorn, He Chen, Yonghui Li, and Branka Vucetic\\}
    \IEEEauthorblockA{        
        School of Electrical and Information Engineering, The University 	of Sydney, Australia \\
        Email: \{matthew.kokshoorn, he.chen, yonghui.li,  branka.vucetic\}@sydney.edu.au }
    \thanks{This work was supported in part by Australian Research Council grants FL160100032, DP150104019 and FT120100487.}
}

\maketitle

\begin{abstract}
This paper develops a novel channel estimation approach for multi-user millimeter wave (mmWave) wireless systems with large antenna arrays. By exploiting the inherent mmWave channel sparsity, we propose a novel simultaneous-estimation with iterative fountain training (SWIFT) framework, in which the average number of channel measurements is adapted to various channel conditions. To this end, the base station (BS) and each user continue to measure the channel with a random subset of transmit/receive beamforming directions until the channel estimate converges. We formulate the channel estimation process as a compressed sensing problem and apply a sparse estimation approach to recover the virtual channel information. As SWIFT does not adapt the BS's transmitting beams to any single user, we are able to estimate all user channels simultaneously. Simulation results show that SWIFT can significantly outperform existing random-beamforming based approaches that use a fixed number of measurements, over a range of signal-to-noise ratios and channel coherence times.
\end{abstract}

\section{Introduction}
	
As microwave frequencies are pushed towards bandwidth-constrained throughput limitations, alternative frequencies are now being considered for 5G cellular systems \cite{niu2015survey}. More specifically, millimeter wave (mmWave) frequencies, ranging from 30GHz to 300GHz, have recently attracted significant attention due to the wide expanse of underutilized bandwidth \cite{heath2016overview}. One fundamental issue of mmWave communications stems from the large free space propagation loss experienced by signals in the high frequency range \cite{rappaportMeasure}. Supplementing this issue, penetration and reflection losses are also much more significant than those at microwave frequencies. As such, the mmWave channel is relatively sparse in the geometric domain, with only a limited number of propagation path directions suitable for conveying information. Overcoming these challenges is now more than ever essential to best utilize the mmWave spectrum, e.g., the 14GHz of the unlicensed spectrum and the 3.85GHz of licensed spectrum recently made available by the FCC in the United States \cite{FFCQual}. 

The most accepted means to overcome and even exploit the inherent mmWave weaknesses, is to implement large antenna arrays so that narrow beams with high beamforming gains can be generated to overcome the severe signal losses \cite{bj2016massive}. Thanks to the small wavelength of the mmWave band, these large arrays can maintain a small form factor. The general idea of mmWave communications is then to steer these narrow beams in the direction of the available propagation paths, effectively ``bouncing'' information-bearing signals off buildings and various other scatters. As a result, in mmWave systems, the sparse channel can estimated by directly finding the beam-steering direction of each path.

Leveraging the sparse characteristic of mmWave geometric channels, previous work has focused on “divide and conquer" type multi-stage algorithms to estimate mmWave channels \cite{rheath,Kokshoorn,7579573,kokshoorn2016race}. These algorithms are essentially path finding algorithms, which divide the process of finding each propagation path into multiple stages. In each subsequent stage, as the user feeds back information to the base station (BS), the estimated angular range is refined so that narrower beam patterns can be used in each following set of channel measurements. These multi-stage approaches have been shown to work well for point-to-point mmWave communications \cite{rheath,Kokshoorn,7579573,kokshoorn2016race}. However, by adapting the BS beam patterns to a specific user, these approaches are inherently limited to estimating only a small number of users in each channel estimation process. As a result, for a multi-user scenario, these types of approaches may no longer be efficient as it could require a training overhead that scales linearly with the number of users. 

Different from these multi-stage adaptive channel estimation algorithms, random beamforming-based approaches are able to carry out simultaneous multi-user channel estimation. Compressed sensing-based channel estimation approaches using random beam-directions and antenna weights have been explored in \cite{mendez2015channel,ramasamy2012compressive,ramasamy2012compressive2,berraki2014application,alkhateeby2015compressed}. These random beamforming-based channel estimation approaches generally perform a predetermined number of random measurements before the channel is estimated.  However, selecting a fixed number of channel measurements does not work well for all users and channel realizations, and may lead to an inferior estimation performance. For example, in a channel realization resulting in a high signal-to-noise ratio (SNR), the channel estimation may not require as many measurements as they would at low SNR. This phenomenon for the multi-user scenario has been discovered in \cite{alkhateeby2015compressed}, wherein different numbers of measurements are required for users with different coherence times and SNRs. However, in reality the multi-user scenario can have users with many different channel characteristics to the BS. As such, it is not feasible to achieve an optimal channel training time that is commonly suitable for all users.

\begin{figure}[!t]
\centering
\includegraphics[width=3.1in,trim={0.5cm 7.4cm 2.0cm 2.1cm},clip]{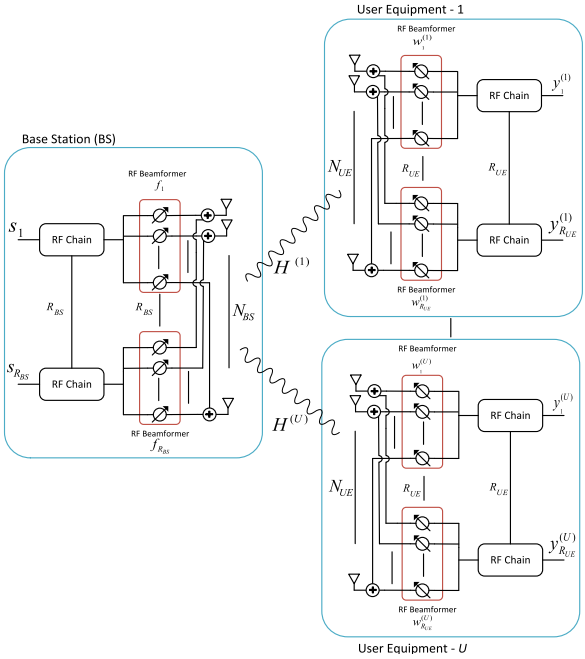}
\caption{System model of the considered multi-user mmWave MIMO system.}
\label{system_model}
\end{figure}

In digital communication systems, this mmWave channel training time adaptation problem is analogous to adapting the transmission rate of communication systems to real-time unknown channel conditions. That is, we also seek to adapt the number of channel estimation measurements, with no prior knowledge of each channel realization.  The rate adaptation problem has led to the development of a powerful rateless coding family known as fountain codes. Inspired by the recently developed concept of analog fountain codes (AFC) \cite{shirvanimoghaddam2013near}, in this paper we develop a novel Simultaneous-estimation With Iterative Fountain Training (SWIFT) framework for the channel estimation of multi-user mmWave MIMO systems. This is achieved by associating the channel estimation problem as equivalent to the AFC design. In SWIFT, the training time required for estimating the multi-user channels is adaptively increased until a predetermined convergence criteria has been met at different users. 

To this end, we propose a ``Fountain code-like'' channel estimation approach, in which the BS keeps transmitting pilot signals in random beam-directions for an indefinite period, essentially encoding random pieces of the channel information into each measurement. At the same time, all users within the BS coverage keep “listening” for these pilot signals by receiving them with random beam-directions. After each measurement, each user estimates its channel based on the pilot signals it has collected and compares it to the previous estimate. If the estimate is similar to the previous estimate (i.e., the estimate has converged), the user regards its channel estimation procedure as complete. The user then feeds back the indices of the BS beamforming vectors to be adopted for its data communication. Simulation results are provided to evaluate the performance of the proposed SWIFT algorithm, which show that SWIFT is able to adaptively adjust its number of channel measurements over a range of SNR values, achieving a superior effective rate when compared to existing schemes using a fixed number of measurements. 

\textit{Notation}: We use letter $\bm{A}$ to denote a matrix, $\bm{a}$ to denote a vector, ${a}$ to denote a scalar, and $\mathcal{A}$ to denote a set. $||\bm{A}||_2$ is the 2-norm of $\bm{A}$. $\bm{A}^T$, $\bm{A}^H$ and $\bm{A}^*$ are the transpose, conjugate transpose and conjugate of $\bm{A}$, respectively. For a square matrix $\bm{A}$, $\bm{A}^{-1}$ represents its inverse. $\bm{I}_N$ is the $N\times N$ identity matrix, $\mathcal{C}\mathcal{N}(\bm{m},\bm{R})$ represents a complex Gaussian random vector with mean $\bm{m}$ and covariance matrix $\bm{R}$.

\section{System Model}
Consider a multi-user mmWave MIMO system comprising of a BS with $N_{\!B\!S}$ antennas and $U$ sets of user equipment (UE), each with $N_{\!U\!E}$ antennas. We consider that the BS and UE are equipped with a limited number of radio frequency (RF) chains, denoted by $R_{\!B\!S}$ and $R_{\!U\!E}$, respectively. To estimate the downlink channel matrix, the BS broadcasts a sequence of beamformed pilot signals to all UEs at the same time. Denote by $\bm{f}_i$ the $N_{\!B\!S} \times 1$ transmitting beamforming vector adopted by the $i$th RF chain at the BS. Similarly, denote by $\bm{w}_j^{(u)}$, the $N_{\!U\!E} \times 1$ receiving beamforming vector adopted by the $j$th RF chain of the $u$th user. In this paper, we consider the beamforming vectors, at each link end, to be limited to networks of RF phase shifters as shown in Fig. \ref{system_model}. As a result, all elements of $\bm{f}_i$ and $\bm{w}_j^{(u)}$ have constant modulus and unit norm such that $||\bm{f}_i||_2=1, \forall$ $i =1,\cdots ,R_{\!B\!S}$, and $||\bm{w}_j^{(u)}||_2=1, \forall$ $j =1,\cdots ,R_{\!U\!E}, u =1,\cdots ,U$. We further consider that each phase shifter (i.e., the entries of $\bm{f}_i$ and $\bm{w}_j^{(u)}$ ) can only use quantized values from a predetermined set given by
\begin{align} \label{BS_set}
\left\{ \frac{1}{\sqrt{N}}\text{exp}( j\pi(-1 + 2(n-1)/N ) \right\}, \forall n=1,...,N,
\end{align}
where $N\in\{N_{\!B\!S},N_{\!U\!E}\}$ is the number of antennas in the array.
%
That is, the BS and UE phase shifters can only use $N_{\!B\!S}$ and $N_{\!U\!E}$ uniformly spaced points around the unit circle, respectively.

We define $\bm{F}=[\bm{f}_1,\bm{f}_2,\cdots,\bm{f}_{R_{\!B\!S}}]$ as the $N_{\!B\!S}\times R_{\!B\!S}$ combined BS beamforming matrix, with columns representing the $R_{\!B\!S}$ RF beamforming vectors. The corresponding $N_{\!B\!S}\times 1$ BS transmit signal can be represented as
\begin{equation} \label{x}
\bm{x} = \sqrt{\frac{P}{R_{\!B\!S}}} \bm{F} \bm{s},
\end{equation}
where $P$ is the total transmit power of the BS and $\boldsymbol{s}$ is the ${R}_{\!B\!S}\times 1$ vector of transmit pilot symbols corresponding to $R_{BS}$ numbers of beamforming vectors with $E[\bm{s}\bm{s}^H]=\bm{I}_{R_{\!B\!S}}$. We adopt a widely-used narrow-band block-fading channel model such that the signal observed by the $u$th user can be expressed as
\begin{align} \label{r}
\bm{r}^{(u)} &= \bm{H}^{(u)}\bm{x} +\bm{q}^{(u)} = \sqrt{\frac{P}{R_{\!B\!S}}} \bm{H}^{(u)}\bm{F} \bm{s} + \bm{q}^{(u)},
\end{align}
where $\bm{H}^{(u)}$ denotes the $N_{\!U\!E}\times N_{\!B\!S}$ MIMO channel matrix between the BS and the $u$th user, and $\bm{q}$ is an $N_{\!U\!E} \times 1$ complex additive white Gaussian noise (AWGN) vector following distribution $\mathcal{C}\mathcal{N}(0, N_0 \bm{I}_{N_{\!U\!E}})$.

Each user processes the received pilot signals with each of the $R_{\!U\!E}$ RF chains. By denoting $\bm{W}^{(u)}=[\bm{w}_1^{(u)},\bm{w}_2^{(u)},\cdots,\bm{w}_{R_{\!U\!E}}^{(u)}]$ as the $N_{\!U\!E}\times R_{\!U\!E}$ combined beamforming matrix at the $u$th user, we express the $R_{\!U\!E} \times 1$ vector of the $u$th user's received signals as
\begin{align} \label{y}
\bm{y}^{(u)} &= (\bm{W}^{(u)})^H \bm{H}^{(u)}\bm{x} +\boldsymbol{n}^{(u)}
\end{align}
where the vector $\boldsymbol{n}^{(u)}=(\bm{W}^{(u)})^H\bm{q}^{(u)} $ follows $\bm{n}^{(u)} \sim \mathcal{C}\mathcal{N}(0, N_0 (\bm{W}^{(u)})^H \bm{W}^{(u)})$.

In this paper, we follow \cite{Sayeed_max} and adopt a two-dimensional (2D) sparse geometric-based channel model. Specifically, we consider that there are $L^{(u)}$ paths between the BS and the $u$th user, with the $u$th user's $l$th path having AOD, $\phi_l^{(u)}$, and AOA, $\theta_l^{(u)}$ with $l=1,...,L^{(u)}$. We further consider these AOD/AOA to be uniformly distributed on the range $[0,2\pi)$. Then the corresponding channel matrix can be expressed in terms of the physical propagation path parameters as
%
\begin{equation} \label{H}
\bm{H}^{(u)} = \sqrt{N_{\!B\!S}N_{\!U\!E}}\sum\limits_{l=1}^{L^{(u)}}\alpha_l^{(u)}   \bm{a}_{\!U\!E}(\theta_l^{(u)}) (\bm{a}_{\!B\!S}(\phi_l^{(u)}))^H
\end{equation}
\noindent
where $\alpha_l^{(u)}\sim\mathcal{C}\mathcal{N}(0,\sigma_R^{(u)})$ is the channel fading coefficient of the $l$th propagation path of the $u$th user, and $\bm{a}_{\!B\!S}(\theta_l^{(u)})$ and $\bm{a}_{\!U\!E}(\phi_l^{(u)})$ respectively denote the BS and UE spatial signatures of the $l$th path. For the purpose of exploration, we consider the BS and each UE to be equipped with linear antenna arrays (ULA). Using ULAs, we can define $\bm{a}_{\!B\!S}(\phi_l^{(u)})= \bm{u}(\phi_l^{(u)},N_{\!B\!S})$ and $\bm{a}_{\!U\!E}(\theta_l^{(u)}) = \bm{u}(\theta_l^{(u)},N_{\!U\!E})$, respectively, where
%
\begin{equation} \label{u}
\bm{u}(\epsilon,N) \triangleq \frac{1}{\sqrt{N}} [1,e^{j \frac{2 \pi d \text{cos}(\epsilon)}{\lambda} },\cdots,e^{j \frac{2\pi d (N-1) \text{cos}(\epsilon)}{\lambda} }]^T.
\end{equation}
\noindent
In (\ref{u}), $N\in\{N_{\!B\!S},N_{\!U\!E}\}$ is the number of antenna elements in the array, $\lambda$ denotes the signal wavelength and $d$ denotes the spacing between antenna elements. With half-wavelength spacing, the distance between antenna elements satisfies $d=\lambda/2$.

To estimate the channel information, at each link end we use beamforming vectors selected from a predetermined set of candidate beamforming vectors. We define the candidate beamfoming matrices as $\bm{F}_c$ and $\bm{W}_c$,  whose columns comprise of all possible candidate beamforming vectors at the BS and UE, respectively. We consider the candidate beams to be the set of all orthogonal beamforming vectors that may later be used for data communication, subject to the quantized phase shifting constraints\footnote{Although we use the hardware limited set of beamforming vectors for ULA, the framework developed in this paper can be used to estimate the gains between any set of more complex candidate beamforming vectors for arbitrary antenna arrays.}. Following (\ref{BS_set}), this leads to $N_{\!B\!S}$ transmitting candidate beams and $N_{\!U\!E}$ combining candidate beams. The $N_{\!U\!E} \times N_{\!B\!S}$ matrix formed by the product of the MIMO channel and these two candidate beamforming matrices is commonly referred to as the virtual channel matrix \cite{rheath} given by
\begin{equation} \label{H_v}
\bm{H}_v^{(u)}= (\bm{W}_c)^H  \bm{H}^{(u)}\bm{F}_c.
\end{equation}

We therefore aim to estimate this matrix so that beam pairs that result in the strongest channel gain can be selected out for data communication. The key challenge here is how to design a sequence of beamforming vectors in such a way that the channel parameters can be quickly and accurately estimated, leaving more time for communication and thus achieving a higher throughput. We assume a block channel fading model with each channel realization having coherence time denoted by $T_c$ symbols. As coherence time is usually quite low for the mmWave frequencies, in the order of hundreds of symbols as used in \cite{alkhateeby2015compressed}, the channel estimation time needs to be kept as short as possible to leave more time for ensuing data communication.

Motivated by the fact that different users may operate in different SNR regions, in next section we develop a fountain code-inspired channel estimation algorithm for the considered multi-user mmWave system, which is able to adapt the number of channel estimation pilot symbols to various channel conditions.

\section{The SWIFT Framework}	

In this section, we first design a set of candidate beamforming vectors to be used in our proposed channel estimation algorithm. We then formulate the channel estimation process as a compressed sensing problem and apply a sparse estimation approach to recover the virtual channel information. Finally, leveraging the introduced beam design and channel recovery scheme, we elaborate the proposed SWIFT framework. 

\subsection{Candidate Beamforming Vectors}


We now design set of candidate beamforming vectors to span the full angular range using quantized phase shifters. To this end, we express the BS candidate beamforming matrix defined in (\ref{H_v}) as $\bm{F}_{c}=[\bm{f}_c(1),...,\bm{f}_c(N_{\!B\!S})]$ and the UE candidate beamforming matrix as $\bm{W}_{c}=[\bm{w}_c(1),...,\bm{w}_c(N_{\!U\!E})]$. Recalling the spatial signatures given in (\ref{u}) and the phase shifting constraints in (\ref{BS_set}), the $n$th BS candidate beamforming vector can then be expressed in terms of the antenna array response vector as
\begin{align} \label{f_c}
\bm{f}_c(n)=\bm{u}\Big(\text{cos}^{-1}\Big(\!\!-\!\!1+\!\!\frac{2(n-1)}{N_{\!B\!S}} \Big),N_{\!B\!S}\Big), \forall n=1,\cdots,N_{\!B\!S}
\end{align}
and the $n$th UE candidate beamforming vector as
\begin{align} \label{w_c}
 \bm{w}_c(n)=\bm{u}\Big(\text{cos}^{-1}\Big(\!\!-\!\!1+\!\!\frac{2(n-1)}{N_{\!U\!E}} \Big),N_{\!U\!E}\Big), \forall n=1,\cdots,N_{\!U\!E}
\end{align}
As the quantized phase shifts are equally spaced around the unit circle, the columns in both candidate beamforming matrices form an orthogonal set and therefore satisfy the properties $ \bm{F}_{c} \bm{F}_{c}^H=\bm{F}_{c}^H \bm{F}_{c}=\bm{I}_{N_{\!B\!S}}$ and  $\bm{W}_{c} \bm{W}_{c}^H=\bm{W}_{c}^H \bm{W}_{c}=\bm{I}_{N_{\!U\!E}}$. That is, $\bm{F}_c$ itself and its conjugate transpose $\bm{F}_c^H$, are each equal to their own inverse. As a result, the noise term in (\ref{y}) becomes $\bm{n}^{(u)} \sim \mathcal{C}\mathcal{N}(0, N_0 (\bm{W}^{(u)})^H \bm{W}^{(u)})=\mathcal{C}\mathcal{N}(0, N_0\bm{I}_{UE})$. In the proposed SWIFT framework, we transmit and receive with random combinations of these candidate beamforming vectors in order to estimate the channels of multiple UEs at the same time.

\subsection{Probabilistic Measurement Beam Selection}

We now can carry out channel measurements by adopting a sequence of randomly selected candidate beamforming vectors at both the BS and UEs. Specifically, in the $m$th measurement timeslot, we propose to form $\bm{F}_m$ by randomly selecting $R_{\!B\!S}$ candidate transmit beamforming vectors from $\bm{F}_c$, i.e. the set of candidate vectors\footnote{Alternatively, a random number of beams may be employed in each measurement timeslot, with a similar to concept to weight set and degree distribution in analog fountain codes \cite{shirvanimoghaddam2013near}. Here, for simplicity, we utilize all RF chains.}. Similarly, to form $\bm{W}_m^{(u)}$ at the $u$th user, we randomly select $R_{\!U\!E}$ candidate receive beamforming vectors from $\bm{W}_c$. Following (\ref{y}), we can then express the $u$th user's received signal in the $m$th measurement timeslot as a $R_{\!U\!E}\times 1$ vector given by
\begin{align} \label{y_m}
\bm{y}_m^{(u)} &=  \sqrt{\frac{P}{R_{\!B\!S}}} (\bm{W}^{(u)}_m)^H  \bm{H}^{(u)}\bm{F}_m \bm{s}_m^{(u)}  + \bm{n}_m^{(u)}.
\end{align}
Using an equal probability of selecting each candidate beam, we can express the probability that the $n$th candidate vector $\bm{f}_c(n)$ is included in $\bm{F}_m$ at the BS as
\begin{align} \label{P_f_n}
\text{Pr}(\bm{f}_c(n)\in \bm{F}_m ) &=  \frac{R_{\!B\!S}}{N_{\!B\!S}}, \forall n=1,...,N_{\!B\!S} 
\end{align}
At each UE we similarly have
\begin{align} \label{P_w_n}
\text{Pr}(\bm{w}_c(n)\in \bm{W}_m^{(u)} )  &= \frac{R_{\!U\!E}}{N_{\!U\!E}}, \forall n=1,...,N_{\!U\!E}.
\end{align}
In all cases, we assume that the BS uses pseudo-random number generator that can therefore be predicted by each user, i.e., each UE knows which random beam selection the BS has made. Note that we have introduced our framework with uniform beam probabilities as described in (\ref{P_f_n}) and (\ref{P_w_n}), and left the optimization of beam selection probabilities as our future work. However, we later show that the proposed scheme works well even for uniform beam selection probabilities.

We conclude this sub-section by expressing the sequence of all observations up to $m$th measurement collected at the $u$th user by a $ m R_{\!U\!E}\times 1$ vector given by

\begin{align} \label{y_all}
\bm{y}^{(u,m)} &=
\left[\begin{array}{c}
\bm{y}_1^{(u)} \\
\vdots   \\
\bm{y}_m^{(u)}
\end{array}\right]\\ &= \sqrt{\frac{P}{R_{\!B\!S}}}
\left[\begin{array}{c}
 (\bm{W}_1^{(u)})^H  \bm{H}^{(u)}\bm{F}_1 \bm{s}_1   \\
\vdots   \\
 (\bm{W}_m^{(u)})^H  \bm{H}^{(u)}\bm{F}_m \bm{s}_m
\end{array}\right] +
\left[\begin{array}{c}
\bm{n}_1^{(u)} \\
\vdots   \\
 \bm{n}_m^{(u)}
\end{array}\right].
\end{align}

\subsection{Sparse Problem Formulation}

In order to recover the channel information using compressed sensing techniques, we require a standard-form expression \cite{vila2011expectation}, $\bm{y}^{(u,m)} = c\bm{A}^{(u,m)} \bm{v}^{(u)} + \bm{n}^{(u,m)}$, where $\bm{A}^{(u,m)}$ is a  $m R_{\!U\!E}\times N_{\!B\!S}N_{\!U\!E}$ sensing matrix, $c$ is some scalar constant, and $\bm{v}^{(u)}=\text{vec}(\bm{H}_v^{(u)})$ is the $N_{\!B\!S}N_{\!U\!E} \times 1$ vectorized virtual channel matrix to be detected.

To this end, we first rearrange (\ref{H_v}) by multiplying it by the left and right hand pseudo inverses of $\bm{W}_c^H$ and $\bm{F}_c$ respectively. We then have
\begin{align} \label{H_H_v}
 \bm{W}_c(\bm{W}_c^H & \bm{W}_c)^{-1}\bm{H}_v^{(u)}  (\bm{F}_c^H \bm{F}_c)^{-1} \bm{F}_c^H  =  \\ & \bm{W}_c(\bm{W}_c^H\bm{W}_c)^{-1}  (\bm{W}_c)^H  \bm{H}^{(u)}\bm{F}_c (\bm{F}_c^H \bm{F}_c)^{-1} \bm{F}_c^H \nonumber
\end{align}
which, after rearrangement, becomes
\begin{align} 
      \bm{H}^{(u)} &= \bm{W}_c\bm{H}_v^{(u)}   \bm{F}_c^H \label{H_H_v_3}
\end{align}
where the simplification follows by the fact that $\bm{W}_c$ and $\bm{F}_c$ are matrices with orthogonal columns leading to $\bm{W}_c^H\bm{W}_c=\bm{I}_{N_{\!U\!E}}$ and $\bm{F}_c^H \bm{F}_c=\bm{I}_{N_{\!B\!S}}$. We can then substitute (\ref{H_H_v_3}) into (\ref{y_m}) to give
\begin{align} \label{y_m_H_v}
\bm{y}_m^{(u)} &=  \sqrt{\frac{P}{R_{\!B\!S}}} (\bm{W}_m^{(u)})^H  \bm{W}_c\bm{H}_v^{(u)}   \bm{F}_c^H \bm{F}_m \bm{s}_m  + \bm{n}_m^{(u)}.
\end{align}
By noticing that $\bm{y}_m^{(u)}$ is already a vector, we can then apply the property $\text{vec}(\bm{A}\bm{B}\bm{C}) = (\bm{C}^T \otimes\bm{A})  \text{vec}(\bm{B})$ to rewrite (\ref{y_m_H_v}) as
\begin{align} \label{y_m_H_v_vec}
\bm{y}_m^{(u)} \!  &= \! \sqrt{\frac{P}{R_{\!B\!S}}} \big(   (\bm{F}_c^H \bm{F}_m \bm{s}_m)^T \!\! \otimes \!    (\bm{W}_m^{(u)})^H  \bm{W}_c  \big)   \text{vec}(\bm{H}_v)     +   \bm{n}_m ^{(u)} \\
          \!  &= \!  \sqrt{\frac{P}{R_{\!B\!S}}}   \bm{A}_m^{(u)}   \text{vec}(\bm{H}_v)     +   \bm{n}_m^{(u)}  \label{y_m_H_v_vec_sub}
\end{align}
where $\bm{A}_m^{(u)}=( \bm{s}_m^T \bm{F}_m^T \bm{F}_c^*)  \otimes   (  (\bm{W}_m^{(u)})^H  \bm{W}_c  ) $ is the $R_{\!U\!E}\times N_{\!B\!S}N_{\!U\!E}$ sensing matrix for the $m$th measurement. Finally, by substituting (\ref{y_m_H_v_vec_sub}) into (\ref{y_all}), we get
\begin{align} \label{y_all_CS}
\bm{y}^{(u,m)} &=  \sqrt{\frac{P}{R_{\!B\!S}}}
\left[\begin{array}{c}
\bm{A}_1^{(u)} \\
\vdots   \\
\bm{A}_m^{(u)}
\end{array}\right] \text{vec}(\bm{H}_v^{(u)}) +
\left[\begin{array}{c}
\bm{n}_1^{(u)} \\
\vdots   \\
 \bm{n}_m^{(u)}
\end{array}\right] \\
&=  \sqrt{\frac{P}{R_{\!B\!S}}}    \bm{A}^{(u,m)} \bm{v}^{(u)} + \bm{n}^{(u,m)}.
\end{align}
%
\begin{figure}[!t]
\centering
\includegraphics[width=3.0in,trim={0.6cm 2.4cm 2.0cm 1.05cm},clip]{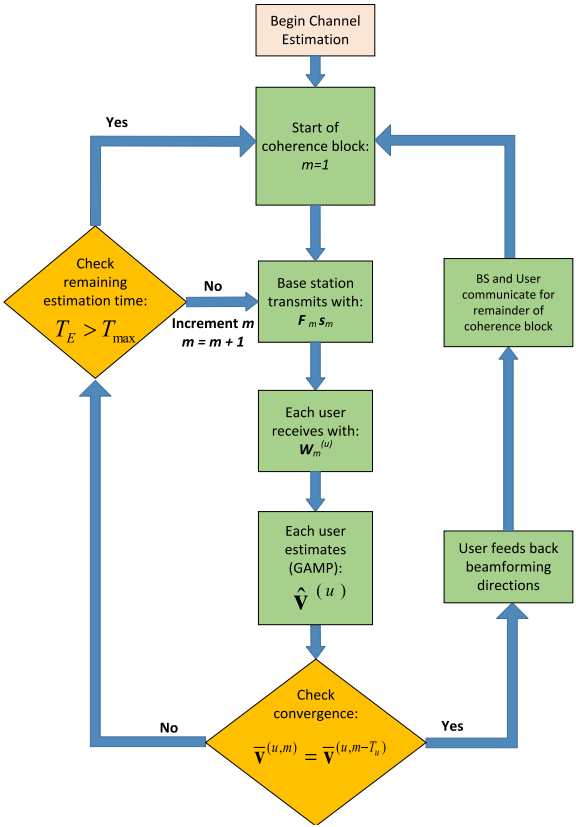}
\caption{Channel estimation flow diagram for each user in the proposed SWIFT framework.}
\label{flow_diagram}
\end{figure}

To complete the problem formulation, we now describe the statistics of the virtual channel vector $\bm{v}^{(u)}$. Although the AOD/AOA are distributed on the continuous ranges in practice and in our simulations, to simplify the estimation problem each UE assumes the AOD/AOA are quantized to those steering directions of the candidate beams given in (\ref{f_c})-(\ref{w_c}). Physically, this is the case where the AOD/AOA are perfectly aligned with each of the candidate beams such that each propagation path can be measured by only one beam combination.

In this case, recalling $\alpha_l^{(u)}\sim\mathcal{C}\mathcal{N}(0,\sigma_R^{(u)})$, the channel sparsity can be characterized by a Bernoulli-Gaussian distribution, in which the $i$th entry of the vectorized virtual channel matrix $\bm{v}^{(u)}$ follows \cite{Jianhua_heath}
\begin{align} \label{v_i_prob}
{v_i^{(u)}} &= \begin{cases}
    	  0 ,& \text{with probability  } 1- \rho \\
    	  \mathcal{C}\mathcal{N}(0,\sigma_R^{(u)}) & \text{with probability  } \rho 	
		 \end{cases} 
\end{align}
for all $i=1,\cdots,N_{\!B\!S}N_{\!U\!E}$ and $\rho= L^{(u)}/(N_{\!B\!S}N_{\!U\!E})$ characterizes the degree of the channel sparsity. With this priori model, we can leverage compressed sensing based sparse estimation methods to recover the channel information. More specifically, we adopt the Bernoulli-Gaussian Generalized Approximate Message Passing (BG-GAMP) estimator\footnote{We omit the details of this estimator due to space limitation. Interested readers are referred to \cite{vila2011expectation}.} as developed in \cite{vila2011expectation} to obtain an estimate of the vectorized virtual channel after $m$th measurement timeslot, denoted by $\hat{\bm{v}}^{(u,m)}$.

\subsection{Stopping Criterion}

As the proposed BS beam patterns do not adapt to any particular user, SWIFT is able to simultaneously estimate all downlink channels for multiple users. To this end, we propose that the BS continues to transmit pilot signals with randomly selected beamfoming vectors, until each user's channel estimation has accurately converged. Recalling (\ref{v_i_prob}), we can assess the channel estimation convergence at the $u$th user by binarizing the estimated virtual channel vector as
\begin{align} \label{v_gamma}
\bar{{v}}_i^{(u,m)} &= \begin{cases}
    	  0 ,& \text{if } |\hat{{v}}_i^{(u,m)}|< \Gamma \sigma_R^{(u)} \\
    	  1 ,& \text{otherwise}	
		 \end{cases}
\end{align}
where $\Gamma<<1$ determines the threshold in which path coefficients are considered negligible or in a deep fade\footnote{In practice, $\Gamma$ may be set according to the minimum fading coefficient to which the transceiver can use for communication, and would depend on the on the required rate of the system, transmit power etc.}. We then consider that the channel estimate has converged if the new binarized virtual channel vector estimate is equal to the previous one, and if there is at least one non-zero element in the vector. That is, the channel estimation of the $u$th user is deemed as complete if $\bar{\bm{v}}^{(u,m)}=\bar{\bm{v}}^{(u,m-T_u)}$ and $||\bar{\bm{v}}^{(u,m)}||_2 \neq 0$, where $T_u$ determines how many measurements are carried out between BG-GAMP estimation updates. We define the time in terms of symbols required for the $u$th user to reach this stopping criterion, $T_E^{(u)}$. To prevent an infinite sequence of measurements when the channel is in a deep fade, we  introduce a maximum allowed number of measurements, denoted by $T_{max}$. Similar limits are also employed in fountain codes to prevent the rate dropping below a certain threshold.

\subsection{Beam Selection for Data Communication}

After meeting the channel estimation stopping criterion, the user stops its estimation process and feeds back the indices of beamforming vectors to be adopted by the BS for the ensuing data communication. To determine these beamforming indices, the user converts the estimated channel vector $\hat{\bm{v}}^{(u,T_E^{(u)})}$ back into its matrix form i.e., $\hat{\bm{{H}}}_v^{ { { (u,T_E^{(u)}) } } }$. The user then determines the candidate beams (for both the BS and UE) that maximize the achievable rate. Recalling the transceiver relationship equations in (\ref{x})-(\ref{y}), this involves finding a BS beamforming matrix, $\bm{F}_d$, and user beamforming matrix, $\bm{W}_d$, that maximizes the achievable rate of the $u$th user given by \cite{rheath} 
\begin{align} \label{Rate}
R_{opt}^{(u)}= \text{log}_2|\bm{I} + \frac{P}{N_0} \bm{W}_d^H \hat{\bm{H}}^{(u,T_E^{(u)})} \bm{F}_d \bm{F}_d^H \hat{\bm{H}}^H \bm{W}_d|.
\end{align}
Recalling from (\ref{H_H_v}) that $\bm{H}^{(u,T_E^{(u)}))}= \bm{W}_c\bm{H}_v^{(u,m)}   \bm{F}_c^H$, we then have

\begin{align} \label{Rate_opt}
\{ \bm{F}_{opt}^{(u)},\bm{W}_{opt}^{(u)} \} &=  \underset{\bm{F}_d,\bm{W}_d}{\operatorname{argmax}}   \text{ log}_2|\bm{I} +  \\  \frac{P}{N_0} \bm{W}_d^H \bm{W}_c&\hat{\bm{H}}_v^{(u,T_E^{(u)})}\bm{F}_c^H \bm{F}_d \bm{F}_d^H \bm{F}_c  (\hat{\bm{H}}_v^{(u,T_E^{(u)})})^H \bm{W}_c^H \bm{W}_d|. \nonumber
\end{align}
As the communication beamforming matrices can only be formed from the candidate beamforming vectors, to resolve (\ref{Rate_opt}), we need to find the BS/UE candidate beams indices to be used for communication. Owing to the mutual orthogonality among the candidate beams, this problem can be reduced to finding the indexes of the dominant values in $\hat{\bm{H}}_v^{(u,T_E^{(u)})}$. Due to the limited feedback bandwidth in the multi-user scenario, we consider that each user is only able to feedback the BS-side beamforming directions determined by (\ref{Rate_opt}), and not the path fading coefficient. As such, we assume that the BS allocates equal power among all identified paths. This reduces the number of feedback bits to only $\lceil\text{log}_2(N_{BS})\rceil$ per estimated path.

We further propose that once a user believes that it has completed its estimation and feeds back the beamforming directions, the BS will use the feedback information and start to communicate with this user using an adjacent sub-channel straight away. The BS can continue to broadcast pilot signals on the previous sub-channel for other users that have not finished their channel estimation. Similar out of band approaches have also been proposed in \cite{nitsche2015steering}. As the relative change in frequency for using an adjacent sub-channel is quite low in the mmWave band, it is reasonable to assume that the AOD/AOA directions remain unchanged in the adjacent sub-carrier, although we acknowledge that in practice a few initial pilots may be required in the new sub-channel to refine the estimate of the fading coefficient. Extension to time and spatial domain multiplexing may also be possible as the BS coordinates the usage of all beamforming directions among multiple users.

To characterize the performance of the proposed SWIFT algorithm, we follow \cite{alkhateeby2015compressed} and define the effective rate of the $u$th user, given the time consumed for the channel estimation, by 
\begin{align} \label{effective_rate}
 R_{E}^{(u)} =  R_{opt}^{(u)} \big(1 - \frac{T_E^{(u)}}{T_c}\big).
\end{align}
recalling that $T_c$ is the coherence time of each channel realization.
\subsection{SWIFT Overview}

We are now ready to overview the proposed SWIFT framework. To this end, we provide an flow diagram of the complete SWIFT channel estimation algorithm in Fig. \ref{flow_diagram}. We also elaborate each step in SWIFT as follows:

\begin{enumerate}
\setlength{\itemindent}{0.18in}
\item[Step (1)]  In each measurement timeslot, the BS randomly selects $R_{BS}$ beamforming vectors to transmit the pilot signals.
\item[Step (2)] At the same time, each user randomly selects $R_{\!U\!E}$ beamforming vectors to receive the pilot signals.
\item[Step (3)] Each user implements the BG-GAMP algorithm to estimate its channel information based on all the collected measurements until the current timeslot.
\item[Step (4)] If the estimated channel has not converged to a the predefined accuracy and the maximum estimation time $T_{max}$ has not been reached, go back to Step 1. Otherwise the channel estimation is considered to be complete and this user can proceed to Step 5.
\item[Step (5)] The user determines the optimal beamforming vectors to be used for communication and feeds back the beamfoming indices for the BS to perform data transmission in the remaining $T_c-T_E$ timeslots.
\end{enumerate}
At the beginning of each transmission block, the process returns to Step 1 and repeats. We end this section by highlighting several of the key benefits of the proposed SWIFT scheme as follows:
\begin{itemize}
\item{Due to the stochastic nature of SWIFT, user feedbacks events are distributed randomly throughout the whole estimation procedure resulting in less pressure on the bandwidth of  feedback channels. }
\item{As our algorithm is inherently designed for various channel estimations with different estimation times, the extension to include a range of different number of antennas and RF chains at the UEs is straightforward.}
\item{As the time occurrence of user feedbacks gives an insight into channel quality, without any additional feedback other than directions of paths. This implicit channel quality information could be leveraged to achieve certain QoS requirements.}
\item{The probabilistic nature of the beam selection naturally allows any prior/partial channel knowledge to be applied to improve channel estimation performance, e.g., a state transition from the previous coherence block's channel.}
\end{itemize}

\begin{figure}[!t]
\centering
\includegraphics[width=3.5in,trim={1.2cm 6.9cm 1.5cm 9.0cm},clip]{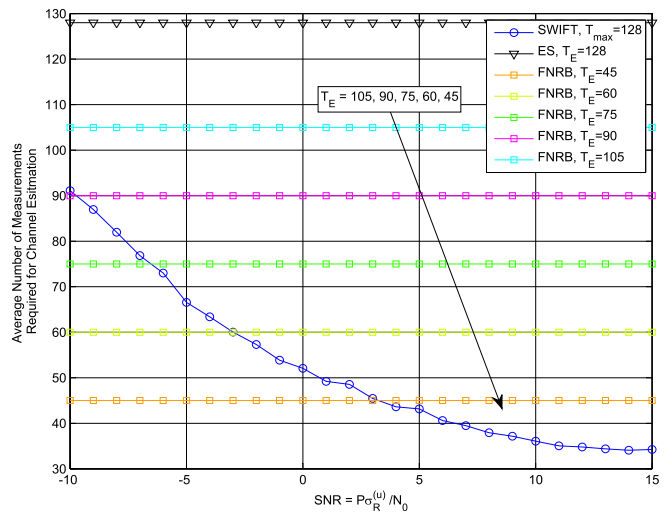}
\caption{Average number of measurements required for channel estimation when the BS equipped with $N_{\!B\!S}=32$ antenna and $R_{\!B\!S}=8$ RF chains and the user is equipped with $N_{\!U\!E}=16$ antenna and $R_{\!U\!E}=4$ RF chains. We assume the number of paths is $L^{(u)}$=1 and update the channel estimate every $T_u=4$ measurements.}
\label{SU_M}
\end{figure}

\begin{figure*}[!h]
\subfigure[][]{\includegraphics[width=3.5in,trim={1.2cm 6.9cm 1.5cm 8.0cm},clip]{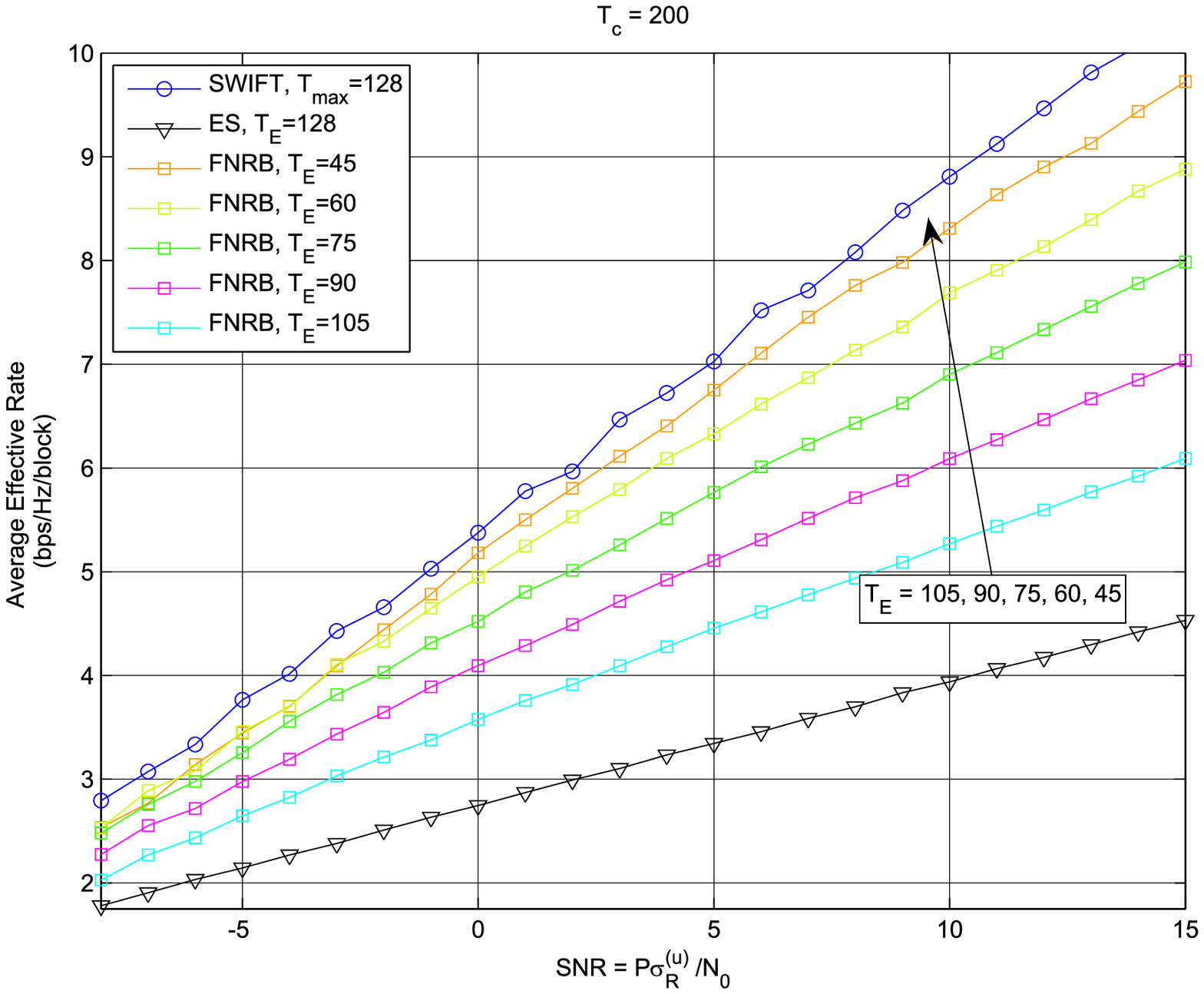}}
\subfigure[][]{\includegraphics[width=3.5in,trim={1.2cm 6.9cm 1.5cm 8.0cm},clip]{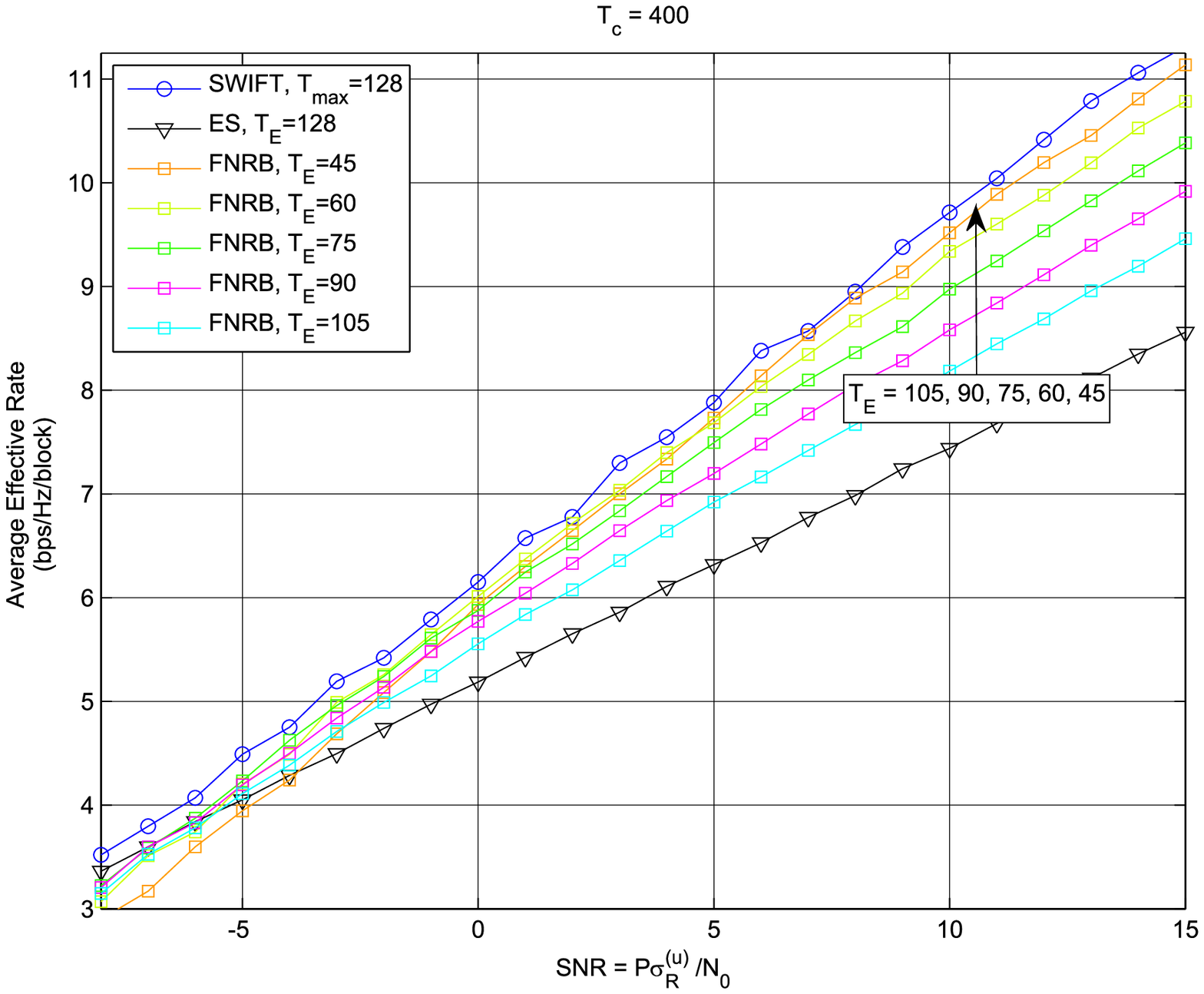}}
\caption{Per-user effective rate for (a) $T_c=200$ and (b) $T_c=400$ when the BS equipped with $N_{\!B\!S}=32$ antenna and $R_{\!B\!S}=8$ RF chains and the user is equipped with $N_{\!U\!E}=16$ antenna and $R_{\!U\!E}=4$ RF chains. We assume the number of paths is $L^{(u)}$=1 and update the channel estimate every $T_u=4$ measurement timeslots.}
\label{SU_rate}
\end{figure*}

\noindent

\section{Numerical Results}

We now provide some numerical results to illustrate the performance of our proposed SWIFT algorithm. We consider a mmWave system with $N_{\!B\!S} = 32$ antennas at the BS and $N_{\!U\!E} = 16$ antennas at each user. We further consider the BS to be equipped with $R_{\!B\!S}=8$ RF chains and each user to be equipped with $R_{\!U\!E}=4$ RF chains. Here, for simplicity, we consider the single path case with $L^{(u)}=1$ with AOD and AOD uniformly distributed on the continuous range $[0,2\pi)$. We also set the maximum allowed number of measurements the same as the exhaustive search-based approach, i.e., $T_{max}=N_{\!B\!S}N_{\!U\!E}/R_{\!U\!E}$. We update the channel estimate every $T_u=N_{\!U\!E}/R_{\!U\!E}=4$ measurements and use $\Gamma=10^{-1}$ in the binarization of the estimated channel vector. 

We compare the proposed algorithm with the benchmark exhaustive search-based approach, in which an estimate of the virtual channel can be found by individually measuring the gains between all combinations of the candidate vectors (i.e., transmitting with only a single beamforming vector but receiving with $R_{\!U\!E}$ beamforming vector(s) in each measurement). We also compare our scheme with those random beamforming-based channel estimation approaches using a predetermined fixed number of measurements, which is represented by FNRB (i.e., fixed number, number beamforming) in all figures. The adopted BG-GAMP estimator used in SWIFT is also applied in the FNRB schemes to estimate the channel information.

Fig. \ref{SU_M} shows the average number of channel measurements required in each of the aforementioned approaches. As can be seen from this figure, the SWIFT algorithm is able to adaptively increase the number of measurements at low SNR values in order to meet the predefined channel estimation accuracy. As all other algorithms use a fixed number of measurements, their average number of measurements remains unchanged across the whole SNR range. 

To best illustrate the performance tradeoff between estimation accuracy and time, Fig. \ref{SU_rate} plots the curves of the average per-user effective rate defined in (\ref{effective_rate}) for various schemes. Specifically, Fig. \ref{SU_rate} (a) and Fig. \ref{SU_rate} (b) show the effective rate for channels with coherence times of  $T_c =200$ and  $T_c =400$ symbols, respectively. From these two subfigures, we can observe that the SWIFT approach can achieve a superior effective rate performance over a very large range of SNR values and different coherence times. In contrast, we see that various schemes using a fixed number of measurements have better performance than one another, depending on both SNR and coherence times. For example, schemes using more measurements perform better at low SNR but worse at high SNR.

\section{Conclusion}

In this paper we have proposed a novel Simultaneous-estimation With Iterative Fountain Training (SWIFT) framework for multi-user channel estimation in mmWave MIMO communication systems. In the proposed algorithm, additional measurements are carried out in an adaptive manner when required, allowing channel estimate to converge to the predetermined accuracy. We have shown that the proposed approach yields superior effective rate performance when compared to those random beamforming-based approaches with fixed number of measurements.

\bibliography{IEEEabrv,\jobname}

\end{document}